\documentclass[%
 reprint,
 amsmath,amssymb,
 aps,
]{revtex4-1}

\usepackage{mathrsfs}
\usepackage{xcolor}
\definecolor{mygreen}{rgb}{0, 0.6, 0}

\def\bea{\begin{eqnarray}}
\def\eea{\end{eqnarray}}

\usepackage{soul,xcolor}
\setstcolor{red}

\usepackage{array}

\usepackage{graphicx}
\usepackage{dcolumn}
\usepackage{bm}

\begin{document}


\title{Rotation-Beating dynamics of a driven flexible filament: role of motor protein properties}

\author{Amir Khosravanizadeh}
\affiliation{Université Paris Cité, CNRS, Institut Jacques Monod, F-75013 Paris, France}
\author{Serge Dmitrieff}
\email{serge.dmitrieff@ijm.fr}
\affiliation{Université Paris Cité, CNRS, Institut Jacques Monod, F-75013 Paris, France}

\date{\today}

\begin{abstract}
We have used numerical simulations to investigate how the properties of motor proteins control the dynamical behavior of a driven flexible filament. The filament is pinned at one end and positioned on top of a patch of anchored motor proteins, a setup commonly referred to as a spiral gliding assay. In nature, there is a variety of motor proteins with different properties. In this study, we have investigated the role of detachment rate, detachment force, stall force, and unloaded speed of motors on the dynamical behavior of the filament. We found that this system generally can show three different regimes: 1) Fluctuation, where the filament undergoes random fluctuations because the motors are unable to bend it. 2) Rotation, in which the filament bends and then moves continuously in one direction. 3) Beating, where the filament's direction of rotation changes over time. We found that the transition between fluctuation and rotation occurs when motors exert a force sufficient to buckle the filament. The threshold force coincides to the second buckling mode  of a filament undergoing a continuously distributed load. Moreover, we showed that when motors near the pining point work close to their stall force, they get stuck and act as a second pin, leading to the beating regime.

\end{abstract}

\pacs{Valid PACS appear here}
\maketitle
\section{Introduction}
In living cells, actin filaments and microtubules are two major cytoskeletal filaments. Actin is a double helical filament with a radius of  a few nanometers and lengths that can reach several micrometers, classifying it as a thin filament. Actin is a semiflexible polymer with a persistence length of up to 17$\mu m$\cite{gittes1993flexural}. Microtubules are hollow rigid rods with a persistence length of a few millimeters\cite{gittes1993flexural,janson2004bending}. The radius of microtubules is approximately 25$nm$, while their lengths can extend to several micrometers \cite{lodish}. Despite their persistence length, both types of filaments can bend on (sub-) micrometric lengthscales, because of external loads, including the activity of motor proteins (hereby referred to as \emph{motors}) \cite{howard2008molecular,kulic2008role}. Motors hydrolyze adenosine tri-phosphate (ATP) to move directionally along the associated filaments ; motors called myosins walk onto F-actin filaments, while kinesins and dyneins are associated to microtubules \cite{howard2001mechanics}. There exists dimeric motors, that can exert forces on two filaments and rearrange them or form a network\cite{sanchez2012spontaneous,strubing2020wrinkling}. One remarkable example is that of cilia, made of 9+2 microtubules doublets linked by minus end directed dynein motor proteins. Owing to the activity of the dyneins, the doublets bend and generate a continued self-organized beating motion\cite{nicastro2006molecular}. In-vitro experiments show that actin bundles with dimeric myosins also exhibit beating behaviour, suggesting a universal property of bundled filaments with motors \cite{pochitaloff2022flagella}. However, single filaments also can exhibit beating in certain conditions \cite{yadav2024wave}.

A common assay to assess motor properties is the \emph{gliding assay}, in which motors grafted to a coverslip can apply tangential forces to microfilaments in the presence of ATP, and slide them across the surface\cite{bohm1997kinesin,howard1989movement,karan2023cooperation,shee2021semiflexible,maloney2011effects}. However, defects on the surface or inactive motors can act as a pinning point and fix one end of the filament \cite{bourdieu1995spiral,winkelmann1995flexibility,bourdieu1995actin}. In this situation, filaments can exhibit a rich variety of dynamical behavior, such as spiraling, swirling, and beating \cite{amos1991bending,bourdieu1995spiral,sekimoto1995symmetry,sumino2012large,schaller2010polar}. 
Early gliding assays were able to measure the buckled shape of the filaments to estimate the average force exerted by motors \cite{bourdieu1995spiral} and continuum elastic models have been applied to determine the shape of the filament as a function of motor forces \cite{fily2020buckling,de2017spontaneous}. 

There are various motors with different characteristic properties such as length, step size, velocity, stall force, binding rate, and unbinding rate \cite{hirokawa1998kinesin,sweeney2018motor}. While some studies have examined the influence of motor density and filament length \cite{yadav2024wave}, the impact of motor properties remains unexplored. For example, the dynamics of the filament can be significantly influenced by how the attachment-detachment rate depends on external forces and time.

\begin{figure}[htb!]
\centering
\includegraphics[width=1\columnwidth]{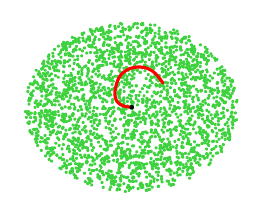}
\caption{Schematic of the system. A thin filament (red) is pined at one end (black dot) positionally but not directionally. Motor proteins in green are anchored to the surface and attempt to walk towards the free (plus) end of the filament. Consequently, they exert a force on the filament, causing it to bend. We observe rotations in either a clockwise or counterclockwise direction. }
\label{fig1}
\end{figure}
In this study, we have used numerical simulations to explore the impact of motor properties on the dynamical behavior of a spiral assay. In section \ref{section2}, we present the details of the numerical model. The results of the simulations are summarized in the section \ref{section3}, where we have found three distinct regimes in the phase diagram of the system: 1) fluctuating regime, where the filament undergoes random fluctuations because motors are not attached to the filament long enough to bend it. 2) Rotation regime, characterized by the bending of the filament as it moves continuously in either a clockwise or counterclockwise direction. 3) Beating regime, in which the filament is moving continuously, but the direction of the filament movement changes over time.

We propose a simple explanation for the phase transitions between these three different regimes, based on motor properties. The fluctuation-rotation transition happens when motors overcome the filament's buckling force, albeit under a continuous load.  Moreover, the motors close to the pinning point are responsible for the transition between the rotation and beating regimes. In this region, if motors operate close to, or above, their stall force, they stall and can act as a second pin, fixing the direction of the filament and leading to its beating.

\section{Model} \label{section2}
The simulations presented in this work are based on a highly coarse-grained description of the cytoskeletal filaments. For this purpose, we use Cytosim\cite{nedelec2007collective,lugo2023typical}, an open-source package based on overdamped Langevin dynamics.

We consider a non-extensible elastic filament represented by $N$ segments and $N+1$ point vertices. Each segment has a fixed length $L/N$, where $L$ is the total length of the filament. The bending energy per unit length of a filament is $\frac{1}{2}\kappa C^2$, where $\kappa$ and $C$ are the bending rigidity and curvature of the filament, respectively. This energy can be written in a discrete form, and forces on the vertices can be derived from it \cite{nedelec2007collective}. The filament is polar, with its minus end fixed at a central position, but the filament still can rotate around the fixed point (see Fig. \ref{fig1}). The filament is placed on top of a substrate of anchored motors, which are randomly distributed within a circular patch of radius $L$. The motors can bind along the filament (not only on vertices) within a binding range $r_b$ at a constant binding rate $\omega_{on}$. Motors are modeled as elastic springs of stiffness $k_m$ ; extension of these springs result in a motor force $\mathbf{f}$ following a hookean law : $\mathbf{f} = - k_m (\mathbf{h} - \mathbf{a})$, with $\mathbf{a}$ the position of the motor anchor and $\mathbf{h}$ the position of the motor head, bound to the filament. 
Bound motors can detach from the filament with an off-rate $\omega_{off}$, that depends on both time and the resistive load,
\begin{equation}\label{Eq2}
\omega_{off}=\omega_d \exp(\| \mathbf{f} \| /f_{d})
\end{equation} 
where $\omega_d$ is a constant detachment rate and $f_{d}$ is the typical unbinding force.

An attached motor tends to move along the filament towards the plus end. The motors implemented in this study are continuous: they do not have a finite step size, but rather move at a velocity $v_m$. This velocity decreases with the projected load $f^T$ against the movement, following the relationship : 
\begin{equation}
v_m=v_0(1-\frac{f^T}{f_s})
\end{equation} 
in which $v_0$ is the motor unloaded speed, $f_s$ is its stall force, and $f^T$ is the projection of the motor force in the local tangential direction $\mathbf{u}$ of the filament towards the minus end :  $f^T = \mathbf{f} \cdot \mathbf{u}$. Thus it is also possible for motors to move towards the minus end if they are being pulled with a tangential force $f^T$ larger than the stall force, $f_s$.

\begin{figure}[t]
\centering
\includegraphics[width=1\columnwidth]{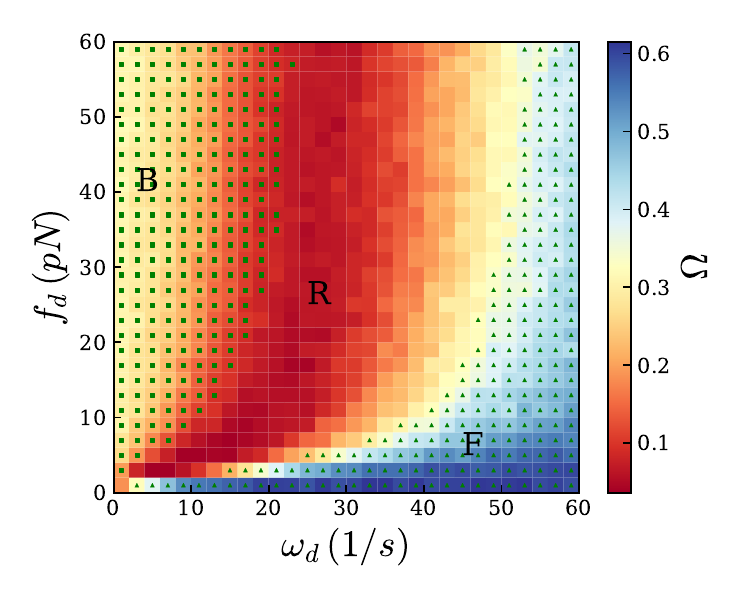}
\caption{Phase diagram where color indicates beating frequency ($\Omega$, denoting the number of changes in the rotation direction of the filament) as a function of detachment rate ($\omega_d$) and detachment force ($f_d$). There are three distinct regimes: fluctuation (F), rotation (R), and beating (b). {\color{green}$\blacksquare$} and {\color{green}$\blacktriangle$} are correspond to the beating and fluctuation regimes, respectively, predicted from our hypotheses. }
\label{fig2}
\end{figure}

\begin{table}[b]
\centering
\begin{tabular}[t]{lll}
\hline
Symbol & Parameter & value \\
\hline
\\
$k_BT$ & Thermal energy &  $4.2 \, pN \cdot nm$ \\
$\Delta t$ &  Time step & $5 \times 10^{-5} \, s$ \\ 
$b_x \times b_y \times b_z$ & Box size & $2 \times 2 \times 0.1\, \mu m^3$ \\
\\
{\bf Filament}\\
\\
L & Filament length & $1 \, \mu m$  \\
$\kappa$ & Bending rigidity & $0.075 \, pN  \mu m^2$ \\
ds & Segmentation length & $10 nm$  \\
\\
{\bf Motors}\\
\\
$N_{MP}$ & number of motors & $1.5 \times 10^{4}$ \\
$\omega_{on}$ & Binding rate & $10 \, s^{-1}$  \\
& Binding range & $5 \, nm$  \\
$f_s$ & Stall force & $5 \, pN$  \\
$v_0$ & Unloaded speed & $2\, \mu m/s$  \\
$k_m$ & Spring stiffness & $500 \, pN / \mu m$  \\

\\
\hline
\end{tabular}
\caption{Simulation parameters.}
\label{tab1}
\end{table}%

\begin{figure*}[t]
\centering
\includegraphics[width=2\columnwidth]{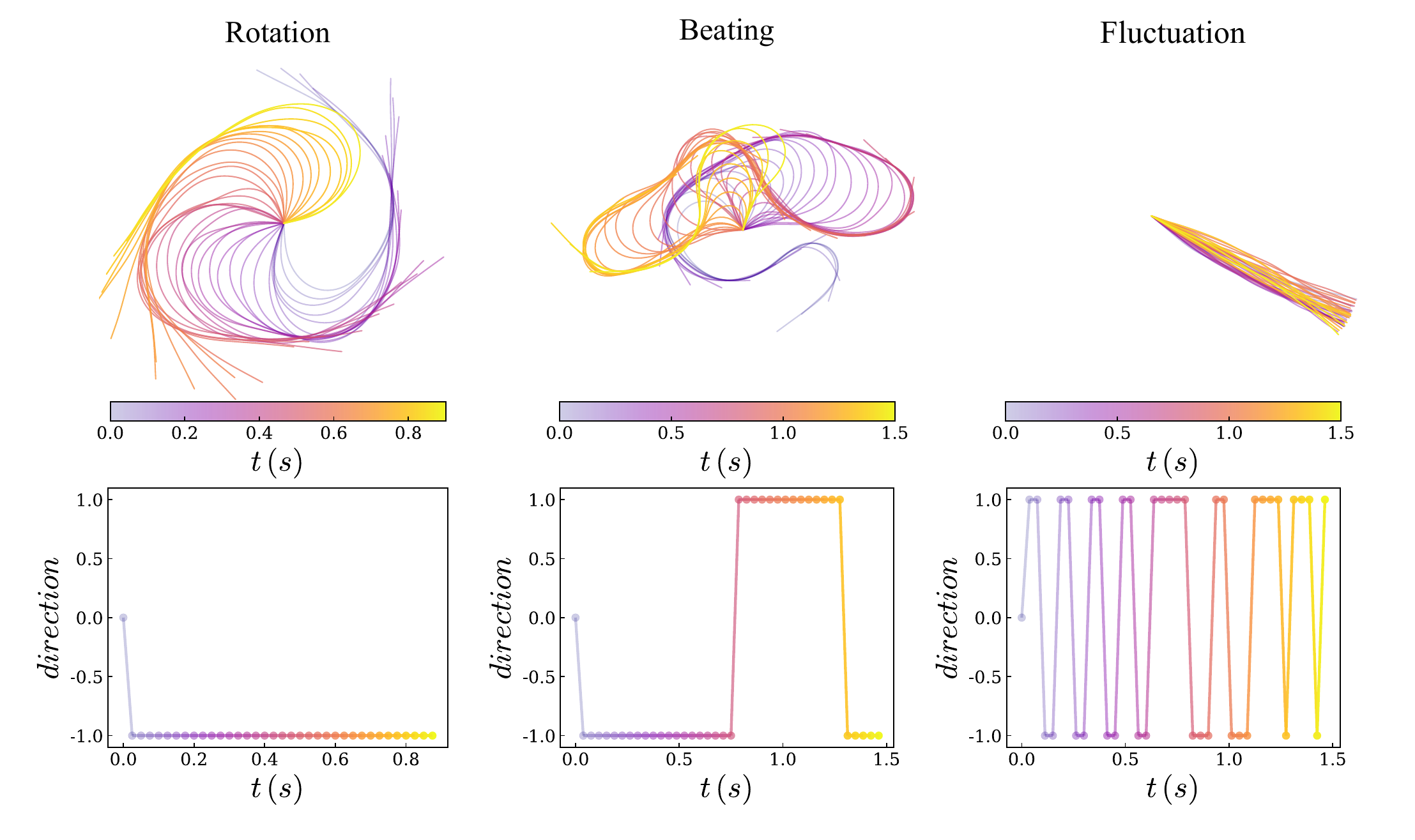}
\caption{Illustration of the three regimes. The top panels illustrate the typical movement of the filament over time, while the bottom graphs display rotation direction as a function of time.  In this representation, $-1$ corresponds to clockwise movement, and $+1$ represents counterclockwise rotation.}
\label{fig3}
\end{figure*}

In this study, we chose to model actin filaments and myosin motors, with parameters chosen in the range of experimental values. All the simulation parameters are summarized in table \ref{tab1}.
We initially run the simulations for $5000$ time steps to reach a steady state. Following that, we continue for $3 \times 10^{5}$ time steps, equivalent to $15$ seconds, while saving frames every $0.15$ seconds.

\section{Results }\label{section3}
To explore the impact of motors properties on the dynamics of a filament with a pinning point at one end, we initially set all parameters as detailed in table \ref{tab1}. Subsequently, by varying the detachment rate ($\omega_d$) and the detachment force ($f_d$), we measure the beating frequency ($\Omega$). The combination of $\omega_d$ and $f_d$ represents the total unbinding rate ($\omega_{off}$) of motor proteins from the filament, as denoted in Eq. \ref{Eq2}. Using the direction of the filament $\mathbf{u}_i$ (at an arbitrary point $i$) over time it is straightforward to calculate the direction of the rotation by the direction of the  vector product $\mathbf{u}_i(t) \times \mathbf{u}_i(t+dt)$. The beating frequency ($\Omega$) is defined as the number of changes in the direction of the rotation divided by the total number of frames (here $100$ frames). If the filament constantly rotates in one direction $\Omega$ is close to zero, and if the direction changes in each frame, it reaches $\Omega=1$. The results of $22500$ simulations for different values of $\omega_d$ and $f_d$ are displayed in Fig. \ref{fig2}, where colors represent $\Omega$. 

\begin{figure*}[t]
\centering
\includegraphics[width=2\columnwidth]{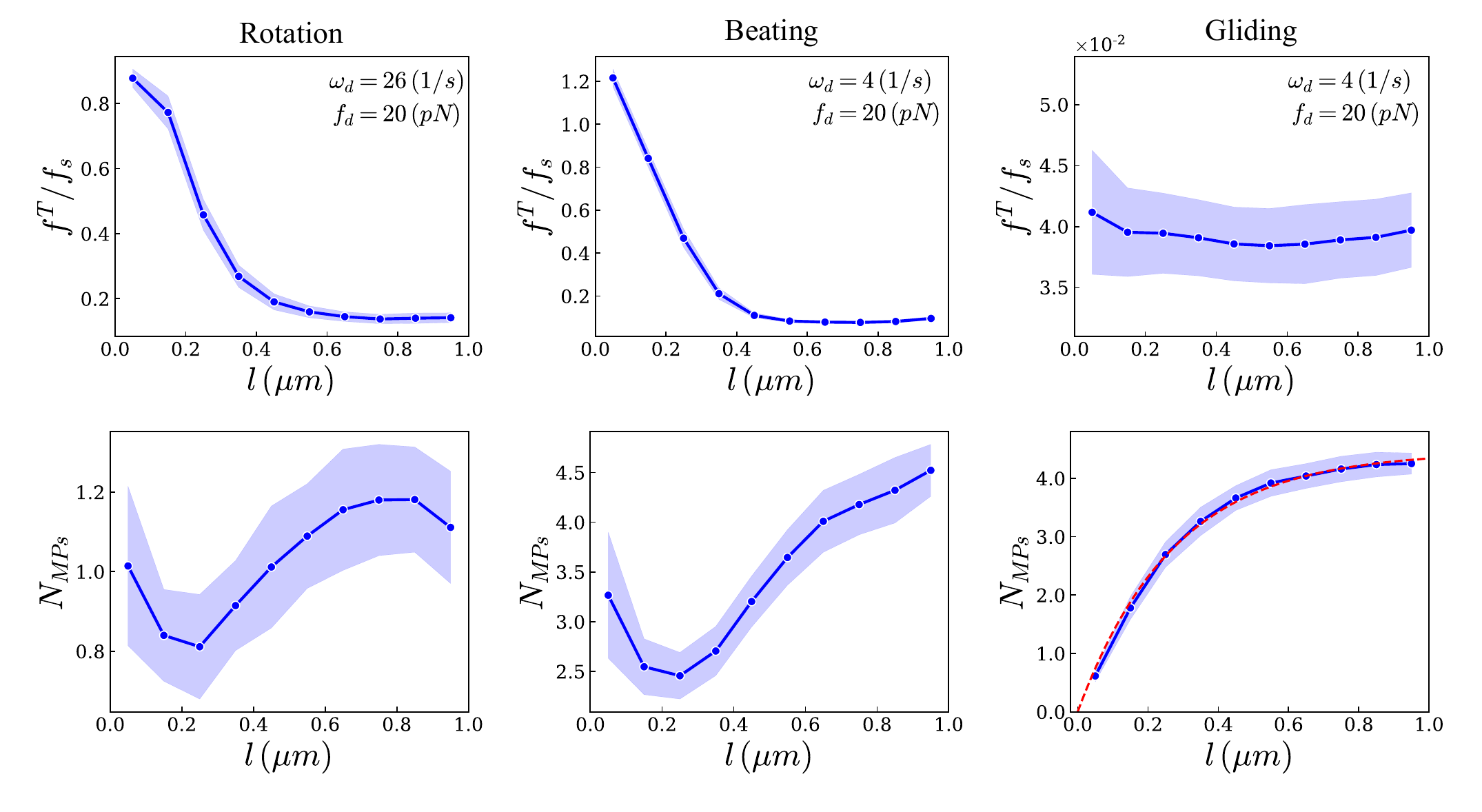}
\caption{At the top, the magnitude of the tangential force is shown as a function of contour length along the filament for the rotation and beating regimes, as well as a gliding assay without any pinning point as a control. In the rotation and gliding regimes, the motors never reach forces higher than the stall force. In contrast, in the beating case, motors near the pinning point can exert forces exceeding their stall force, effectively acting as a second pin. At the bottom, the number of motors is shown as a function of contour length along the filament. The filament is divided into ten equal segments, and both the force and the number of motors are averaged over time for each segment. To minimize stochastic effects, the plots are averaged over a hundred simulations.}
\label{fig4}
\end{figure*}

To reduce the stochasticity of the individual results in the total phase diagram, we averaged individual simulation results with neighboring simulations (close in the $f_d, \omega_d$ parameter space). The original data of individual simulations is presented in supplementary Fig. \ref{figS1}. Because the simulated filaments are not chiral, there is no favorite direction of motion, see supplementary Fig. \ref{figS2}. The phase diagram shows three different regimes, Fig. \ref{fig3} and supplementary movie S.1. For a large detachment rate $\omega_d \gg 1$, or a very small detachment force $f_d \simeq 0$ motors cannot stay attached to the filament long enough ($\omega_{off} \gg 1$) to apply a sufficient force to the filament to bend it. This regime is named 'fluctuation' ($F$) where the filament merely jiggles while remaining mostly straight. Conversely, for small $\omega_d$ and large $f_d$, the system tends to continuously 'beat' ($B$), wherein it rotates in one direction and then changes to another. In between, the system exhibits a continuous 'rotation' ($R$), with few or no change of direction. We also found that the rotation speed changes slowly between the rotation and fluctuation regimes, while it is much lower in the fluctuation regime, Fig. \ref{fig5} and Supplementary Fig. \ref{figS3}.  

\subsection*{Fluctuation to rotation}
We hypothesized that the transition from the fluctuation regime to rotation happens when motor proteins can exert enough force to overcome the buckling force of the filament. Indeed, we could observe a continuous increase of the number of bound motors between the fluctuation and the rotation regime, supplementary Fig. \ref{figS4}, corresponding to an increase of the filament bending energy, supplementary Fig. \ref{figS5}. Here, the force is applied by motors all over filament length, rather than on the ends only, and the buckling threshold should differ from the canonical Euler buckling threshold $f_B = \pi^2 \kappa / L^2$. To model the motor distribution, we assumed a force density $w$. Up to buckling, the filament is straight, and we discard feedback between filament mechanics and motor force, and thus we can assume $w$ to be constant. Similar to  Euler buckling, we can solve the Euler-Lagrange equations, and find the filament shape, see Supporting Information. The buckling threshold is found by computing the first modes that satisfy the boundary conditions. The main difference with Euler buckling is that the force on the free end is $0$, so we can compute the threshold force at the pinned end. We find that the critical force at the pinned end is $f_j^* = \bar{w}_j \frac{ \kappa}{ L^2 }$, with $\bar{w}_j$ the critical force density of mode $j$ : $\bar{w} \approx \left[ 25.64, 95.95, 210.68, 369.828, .... \right]$; note that this force is at least $\sim 2.5 \times$ larger than $f_B$. We should expect the $F \rightarrow R$ transition to occur when the sum of motor forces along the filament is larger than $f_1^*$.
 
However, motors also restrict movement normal to the filament. If motor distribution were continuous, this would result in adding an elastic potential normal to the filament axis.  There is no straightforward solution to the buckling problem with a continuous force along the filament axis and a quadratic potential in the normal axis, but it is clear that the normal potential could strongly increase the buckling force, with or without selecting higher modes \cite{dmitrieff2017balance}. With discrete motors, the distribution of motors along the filament could select higher modes of deformation for filaments \cite{brangwynne2006microtubules}, and thus increase the critical load density. We measured the total force exerted by motors on the filament in simulations, and we found that the buckling threshold coincides with the total motor force being larger than $f_2^*$, the second deformation mode of a filament under a constant load density : in Fig. \ref{fig2} triangles represent the area where the total force is smaller than $f_2^*$. 

\subsection*{Rotation to beating}
The second transition is between the rotation and beating regimes, that both have been observed in experiments. Theoretical work showed that pinning the filament via a second anchor leads to beating\cite{bourdieu1995spiral,fily2020buckling,yadav2024wave}. It was claimed that, in experiments exhibiting beating, this behavior occurs when there is a second defect close to the pining point or the pining point extend to a line\cite{bourdieu1995spiral,bourdieu1995actin}. In our system, there are neither secondary anchors, nor inactive motors, and the beating was a surprise.

We hypothesized that the motors near the pinning point are responsible for the beating of the filament. If these motors operate close to their stall force, they can fix the filament direction, leading to beating.
\begin{figure}[ht!]
\centering
\includegraphics[width=0.8\columnwidth]{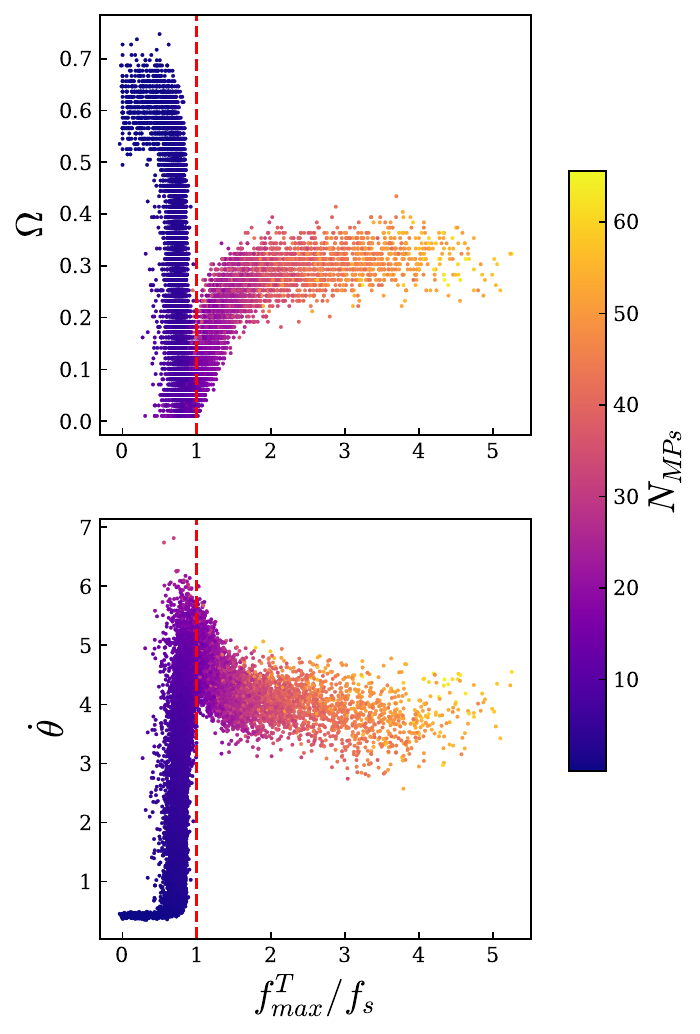}
\caption{The beating frequency (top) and the speed of rotation (bottom) of the filament as a function of the maximum tangential force. A phase transition between the rotation and beating regimes happens around $f^T_{max}=f_s$. The color bar denotes the average number of bound motors.}
\label{fig5}
\end{figure}
To validate this hypothesis, we computed the tangential force, averaged over time, as a function of the arclength. We averaged the mean tangential force over a hundred simulations to reduce the stochasticity of the results.

First, as a control, we computed this tangential force in an unpinned gliding assay, in which the filament is free to slide on a surface of anchored motors. We were able to formulate a master equation to predict the motor density in such a gliding assay, see Supporting Information. Motors were assumed to bind all along the filament, and move along the filament. Because the filament is not pinned, we could assume the motor force to be independent of position ; therefore, the motor velocity and unbinding rate should be independent of position. We therefore predicted an exponential motor distribution as a function of arclength, Fig. \ref{fig4}, right. In simulations, we do find the force per motor to be near-independent of position ; accordingly, we find a near-exponential distribution for the motor density. We also find that the force exerted by motors is only a fraction of the motor stall force : if the filament drag is low enough, motor operates near their maximum speed, at low forces.

In contrast, for a pinned filament, we find a non-monotonous motor density as a function of arclength, Fig. \ref{fig4}, left and middle. While we expect motor binding to be independent of position, forces applied by motors depend on their positions along the filament (Fig. \ref{fig4}, lower panels). Therefore the motor velocity and unbinding rate should depend on position, yielding a coupling between filament mechanics and motor dynamics. In the rotating case (Fig. \ref{fig4}, left), the force exerted by motors is higher than for a gliding assay, because motors have to exert more force to bend the filament than to merely cause a translational motion. The force is higher close the pinned end of the filament, but remains below the stall force.

In contrast, in the beating case, the tangential force near the pinned end  is higher than the stall force. Since an isolated motor cannot exert a force beyond its stall force, this is due to the other motors on the filament deforming the filament, and thus stretching the motors near the pinned end at forces higher than the stall force.  As a consequence, motors close to the pinned (minus) end are being dragged towards the pinned end rather than walking towards the free (plus) end ; this results an accumulation of motors at the pinned end, that are moving in reverse. We believe that this accumulation of motors beyond their stall force is causing the beating.  Individual simulations confirm this accumulation in the beating case (Supplementary Fig. \ref{figS6}) and its absence in the rotation case (Supplementary Fig. \ref{figS7}).

\begin{figure}[ht!]
\centering
\includegraphics[width=1\columnwidth]{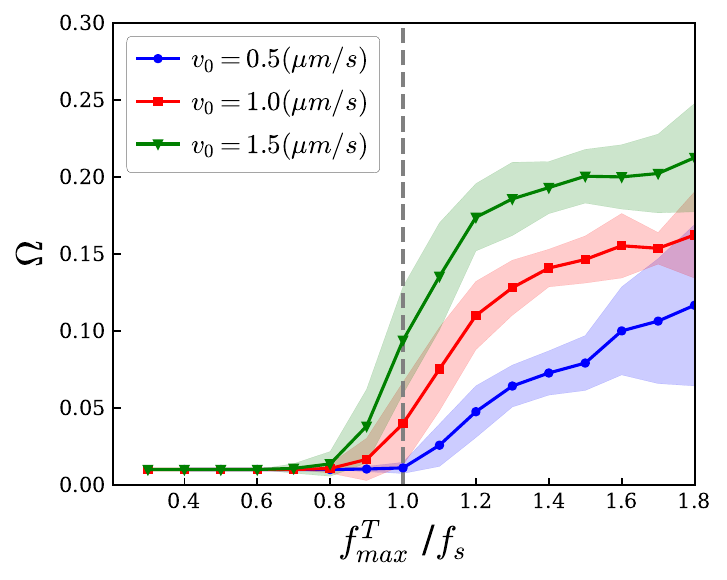}
\caption{The beating frequency as a function of the maximum tangential force for three different unloaded speeds, $v_0=0.5,1.0,1.5 \, (\mu m/s)$. A phase transition between the rotation and beating regimes occurs around $f^T_{max}=f_s$.}
\label{fig7}
\end{figure}
Accordingly, a transition from rotation to beating should occur when the tangential force along the filament surpasses the stall force. In simulations, we computed the maximal tangential force along the filament, averaged over time $f^T_{max}$ ; we found that simulations with $f^T_{max}>f_s$ mostly belonged to the beating phase rather than the rotation space, Fig. \ref{fig2}. Using these simulations results, and plotting the beating frequency of the system as a function of $f^T_{max}$ indeed revealed a phase transition occurring around $f^T_{max}=f_s$, Fig. \ref{fig5}.

To confirm that the rotation to beating transition was caused by motors near the pinned end acting as additional anchors, as the filament is being moved by the other motors on the filaments, we decided to explore the behavior of the system as a function of two other motors parameters. 
We systematically varied the motor stall force $f_s$, for several values of the unloaded speed $v_0$. We found that $\Omega$ increases drastically above a threshold of $f^T_{max}/f_s$ close to $1$, indicating the rotation to beating transition, Fig. \ref{fig7}. The precise threshold of $f^T_{max}/f_s$ is shifted $\approx 20\%$ while $v_0$ is changed by $\approx 300\%$. Therefore, while other parameters play a minor role, the rotation to beating transition is mostly driven by $f^T_{max}$ being larger than $f_s$.

\section{Discussion} \label{Conclusion}
We found three different regimes  for the spiral gliding assay. 1) The fluctuation ($F$) regime, where motors do not exert enough force to bend the filament, that the filament jiggles around its position. 2) The rotation regime($R$), characterized by the filament bending and continuously rotating in one direction. 3) The beating regime ($B$), wherein the filament rotates but the direction of rotation changes over time, alternating between clockwise and counterclockwise.

We found that the $F\rightarrow R$ transition  is a consequence of the motors collectively exerting a force higher than the buckling force of the filament with a transition for a total force larger than $f_2^*  \sim  95.95 \kappa / L^2$. This is approximatively $10x$ larger than the Euler buckling threshold $f_B = \pi^2 \kappa /L^2$.  The buckling threshold $f_2^*$ coincides with the second buckling mode of a pinned filament driven by a constant motor density, that we were able to compute analytically. However, this high buckling force could also stem from the confinement by motors of the filament along its normal direction.

Surprisingly, we also find a $R\rightarrow B$ transition, from rotation to beating, although the filaments have a single pinning point. We proposed that the $R\rightarrow B$ transition results from the motors near the pinning point, that effectively act as a second pin that fixes the local filament direction, leading to its beating as predicted for a clamped filament. 

Our simulations confirmed this hypothesis. These motors at the pinned end operate close to, or above, their stall force, moving slowly or even backwards on the filament, thus accumulating at the pinned end and clamping the filament. This was confirmed by changing the motor stall force, showing that the $R\rightarrow B$ transition is mostly controlled by the $f_{max}^T / f_s$ ratio ; above a threshold value close to $1$, the beating frequency strongly increased in a stepwise manner. 

In a gliding assay without any pinning point, motors never exceed their stall force, while in a spiral system, motors near the pinning point can reach forces even higher than their stall force. This highlights an unexpected coupling between filament mechanics and collective motor action. Accordingly, both transitions are strongly controlled by the motor properties, both motor unbinding rate $w_d$, unbinding force $f_d$, stall force $f_s$ and unloaded velocity $v_0$, in a non-trivial way.

The $R \rightarrow B$ transition was not identified in previous theoretical works, where instead a clamped end (without orientational degree of freedom) was used. However, these works did not include the force dependence of motor velocity, nor the discrete nature of motors \cite{fily2020buckling}, that we found here to play a crucial role.

While gliding assays of single filaments are an over-simplification of biological instances of the cytoskeleton, the strong control by motor properties reveals a crucial property of cytoskeletal networks. While the existence of filament rotation and beating phases are universal consequences of filament-motor assemblies, the phase in which the system lives depends strongly on motor properties, and thus on which precise motor proteins exists locally. Therefore, local signaling pathways (such as Rho, Rac, Cdc42 for actin \cite{byrne2016bistability})
 allow the cell to explore various areas of the phase diagram by recruiting a few specific motors and cytoskeleton-associated molecules.

\section{Acknowledgements}
The authors would like to thank Nicolas Minc for scientific discussions, and François Nédélec for the discussions and critical reading of the manuscript. Financial support from INSERM grant Aviesan "MMINOS" is gratefully acknowledged. We gratefully acknowledge Joel Marchand for the IT support and the iPOP-UP computing cluster.

%

\clearpage

\section{Supporting information}
 \renewcommand\thefigure{S.\arabic{figure}} 
 \setcounter{figure}{0}

\begin{figure}[ht!]
\centering
\includegraphics[width=0.8\columnwidth]{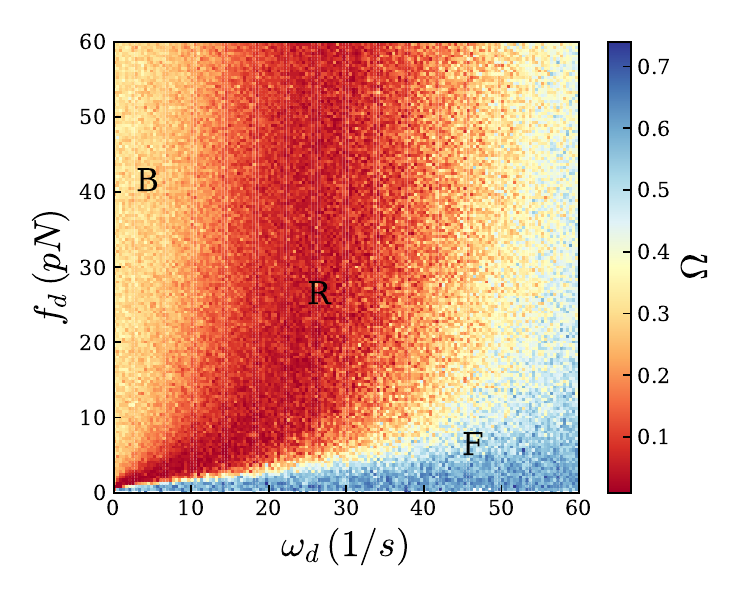}
\caption{Phase diagram of the system for individual simulations. Each dot in this plot represents the average beating frequency of one simulation over time.}
\label{figS1}
\end{figure}

\begin{figure}[ht!]
\centering
\includegraphics[width=0.8\columnwidth]{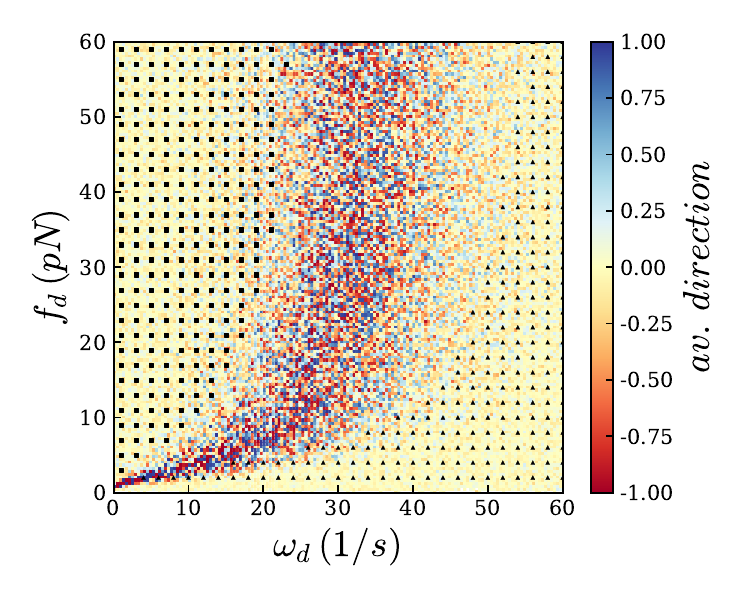}
\caption{Average filament direction  over time as a function of $\omega_d$ and $f_d$.}
\label{figS2}
\end{figure}

\begin{figure}[ht!]
\centering
\includegraphics[width=0.8\columnwidth]{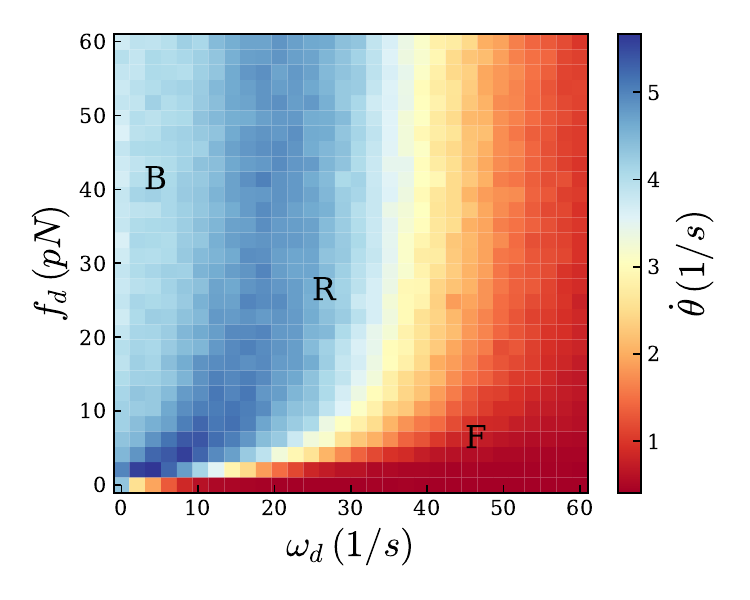}
\caption{Average absolute filament rotation speed as a function of $\omega_d$ and $f_d$.}
\label{figS3}
\end{figure}

\begin{figure}[ht!]
\centering
\vspace{-1.cm}
\includegraphics[width=0.8\columnwidth]{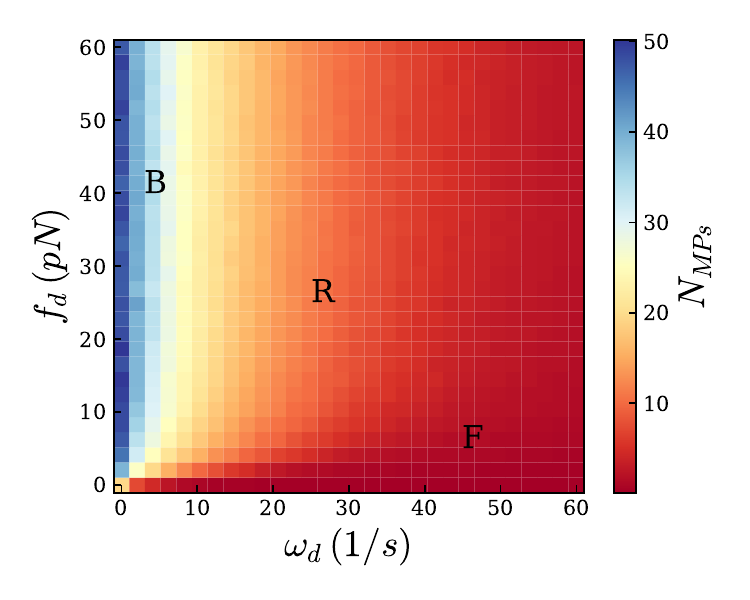}
\caption{Average number of bound motors to the filament as a function of $\omega_d$ and $f_d$.}
\label{figS4}
\end{figure}

\begin{figure}[ht!]
\centering
\includegraphics[width=0.8\columnwidth]{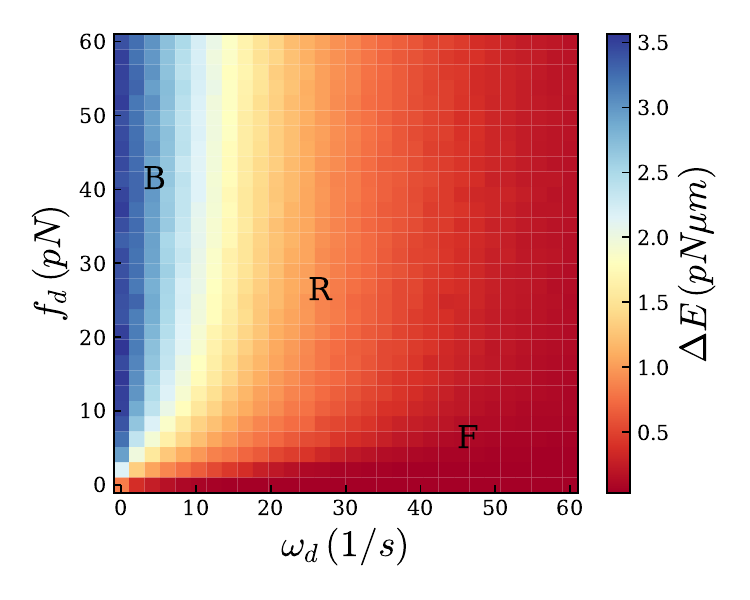}
\caption{Average bending energy of the filament as a function of $\omega_d$ and $f_d$.}
\label{figS5}
\end{figure}

\begin{figure}[ht!]
\centering
\includegraphics[width=0.8\columnwidth]{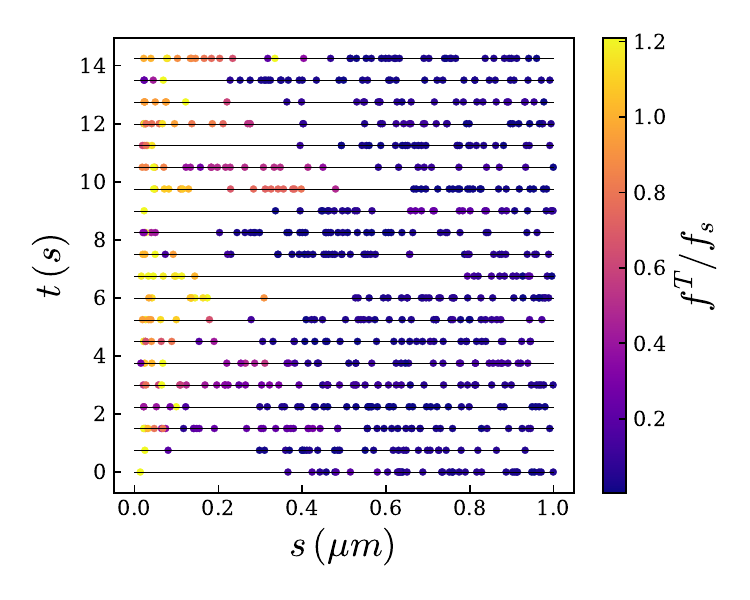}
\caption{Time lapse of motors as a function of the filament's contour length for a beating case. Colors display the tangential force, $f^T/f_s$.}
\label{figS6}
\end{figure}

\begin{figure}[ht!]
\centering
\includegraphics[width=0.8\columnwidth]{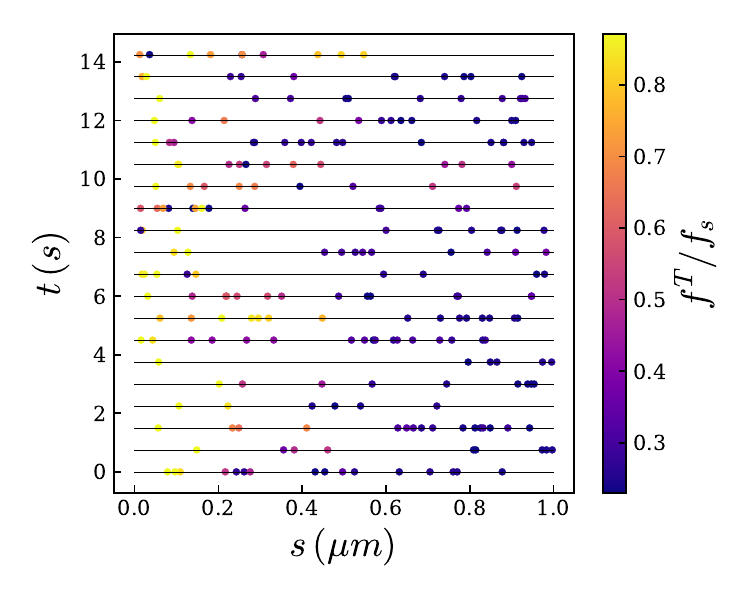}
\caption{Time lapse of motors as a function of the filament's contour length for a rotation case. Colors display the tangential force, $f^T/f_s$.}
\label{figS7}
\end{figure}

\clearpage
\subsection*{Theoretical calculation of motor density along the filament in a gliding assay}

\renewcommand\theequation{S.\arabic{equation}} 
\setcounter{equation}{0}
In a gliding assay, if the length of the filament is smaller than its persistence length, $L\ll L_p$, it can be considered as a straight filament. In this situation, the number of bound motors per unit length of the filament $\rho(s,t)$ satisfies the reaction convection equation 
\begin{equation} \label{eq_S1}
\partial_t \rho(s,t) = - \partial_s J + \rho_0 \, \omega_{on} - \rho(s,t) \, \omega_{off}
\end{equation} 
where $J=\rho \, v$ is the current of motors toward the filament plus end. The binding rate of the motors to the filament is $\rho_0 \, \omega_{on}$ where $\rho_0$ is the density of anchored motors. The bounded motors can detach from the filament with a rate of $\rho(s,t) \, \omega_{off}$. At steady state, motor density constant in time $\partial_t \rho(s)=0$, and equation \ref{eq_S1} can be written as
\begin{equation}
\frac{d \rho(s)  v}{ds} = \rho_0 \, \omega_{on} - \rho \, \omega_{off}, 
\end{equation} 
which can be solved to find $\rho$ as a function of $s$, the contour length of the filament,
\begin{equation}
\rho(s) = \frac{\rho_0 \, \omega_{on}}{\omega_{off}} -A \exp( -\frac{s \omega_{off}}{v}).
\end{equation} 
The boundary condition for this equation is the absence of motors at the minus end of the filament, $\rho(0)=0$. This yields  
\begin{equation}
A=\frac{\rho_0 \, \omega_{on}}{\omega_{off}},
\end{equation} 
and the density of motors can be found as
\begin{equation}
\rho(s) = \frac{\rho_0 \, \omega_{on}}{\omega_{off}} \left( 1- \exp( -\frac{s \omega_{off}}{v}) \right).
\end{equation} 
Then, the average total number of bound motors to the filament can be calculated by integrating $\rho(s)$ over the total length of the filament: 
\begin{eqnarray}
<N_{MPs}>=\int_0^L \rho(s) ds = \qquad \qquad \qquad  \\
\frac{\rho_0 \, \omega_{on}}{\omega_{off}} \left( L-\frac{v}{\omega_{off}} \left( 1- \exp( -\frac{L \omega_{off}}{v}) \right)  \right), \nonumber
\end{eqnarray}
which is an exponential function, and fits very well to the simulation results for a gliding system in Fig. \ref{fig4}.

\subsection*{Buckling of a beam under continuous load}
We describe a beam embedded in 2D space, the coordinates of which can be written $y(x)$. For small deformations, we can write the energy :
\begin{eqnarray}
\mathcal{F} = \frac{1}{2}\int_0^L \kappa \left( y'' \right)^2  - \sigma \left( y' \right)^2 ds
\end{eqnarray}
Euler-Lagrange equation for this system is : $\frac{\partial \mathcal{L}}{\partial y} - \partial_x \frac{\partial \mathcal{L}}{\partial y'} + \partial_x^2 \frac{\partial \mathcal{L}}{\partial y''}=0$, with $\mathcal{L} = \left(  \kappa \left( y'' \right)^2   + \sigma \left( y' \right)^2 \right) /2$. This yields :

\begin{eqnarray}
\partial_x (\sigma y') + \kappa y'''' = 0
\end{eqnarray}
Which can be integrated with respect to $x$ : 
\begin{eqnarray}
\sigma y' + \kappa y''' = A
\end{eqnarray}
In which $A$ is a constant.

Additionally, when deriving the Euler-Lagrange equations, we get boundary conditions :
\begin{eqnarray}
\left[ \delta y' \frac{\partial \mathcal{L}}{\partial y}  \right]_0^L = 0 \\
\left[ \delta y \left( \frac{\partial \mathcal{L}}{\partial y'}  - \partial_x \frac{\partial \mathcal{L}}{\partial y''} \right) \right]_0^L = 0 
\end{eqnarray}
The first condition is always zero, the second  boundary conditions imposes : $- \sigma y' - \kappa y''' = 0$ at the boundaries. Therefore, the shape equations for the buckled beam are :  
\begin{eqnarray}
\sigma y' + \kappa y''' = 0 \label{resultEL}
\end{eqnarray}
Additionally, torque-free conditions at the boundary are : $y''(0)=y''(L)=0$.

Here $\sigma$ is the (1D) pressure in the beam. In classical Euler buckling, $\sigma$ is a constant, and the first solution to Eqn. \ref{resultEL} is $\cos{ x/\sqrt{\kappa / \sigma}} $ with $\sigma = \kappa \pi^2 / L^2$ the critical buckling load.

In our case, $\sigma$ depends on $x$ as it results from the motor distribution. For simplicity, we will assume here a continuous motor force distribution and no additional load. We thus have $\sigma = w (L-x)$, in which $w $ is a force per unit length, that could be written : $w=\rho f_M$, with $f_M$ the motor force and $\rho$ the linear motor density along the filament.

We can normalize equation \ref{resultEL} for simplicity, with $\bar{w} = L^3 w / \kappa$. Moreover, we will write this as an equation for the angles $\theta$, using the small angle approximation $y'=\sin{\theta} \approx \theta$ :  
\begin{eqnarray}
\bar{w}(1-\bar{x}) \theta + \ddot{\theta} = 0 \label{resultEL_theta} 
\end{eqnarray}
In which the $\dot{\theta}$ indicates the spatial derivative with respect to the normalized length $\bar{x}$. Assuming $\dot{\theta}(0)=0$ (torque free condition), we find, up to a constant :
\begin{eqnarray}
\theta(x) \propto  \text{Ai}\left(\sqrt[3]{\bar{w}} (x-1)\right)  \qquad \qquad  \qquad\nonumber \\ - \frac{\text{Ai}'\left(-\sqrt[3]{\bar{w}}\right) \text{Bi}\left(\sqrt[3]{\bar{w}} (x-1)\right)}{\text{Bi}'\left(-\sqrt[3]{\bar{w}}\right)}
\end{eqnarray}
In which $\text{Ai}$ and $\text{Bi}$ are the corresponding Airy functions. From there, we find :
\begin{eqnarray}
\lim_{x\rightarrow 1} \dot{\theta}(x) \propto \sqrt[3]{\bar{w}} \qquad \qquad \qquad \qquad \qquad \qquad \qquad \\
\times \left(-\frac{\sqrt[6]{3} \text{Ai}'\left(-\sqrt[3]{\bar{w}}\right)}{\Gamma \left(\frac{1}{3}\right) \text{Bi}'\left(-\sqrt[3]{\bar{w}}\right)}-\frac{1}{\sqrt[3]{3} \Gamma \left(\frac{1}{3}\right)}\right) \nonumber
\end{eqnarray}
With $\Gamma$ the gamma function. Remember that we need $\dot{\theta}(1)=0$ to satisfy the torque free condition. This means that $\bar{w}$ has to satisfy $\lim_{x\rightarrow 1} \dot{\theta}(x)=0$. While we cannot find analytical solutions to this equation, we can identify numerically the first solutions $\bar{w}_i \approx \left[ 25.64, 95.95, 210.68, 369.828, .... \right]$. 

This means that the critical force density for buckling of a beam under continuous load under mode $i$ is $\rho^* = \bar{w}_i \kappa / L^3 $.  Since $\rho L$ is the load at the pinned boundary ($x=0$), we can predict the critical force at the pinned end : 
\begin{eqnarray}
f_i^* = \bar{w}_i \frac{ \kappa}{ L^2 }
\end{eqnarray}
For $i=1$ (resp. $i=2$) this is around 2.6 times (resp. $9.7 \times$) larger than the Euler buckling threshold $f_B = \pi^2 \kappa / L^2$.

\subsection*{Comparison to simulations}
In simulations, we find that the fluctuations-rotation transition ($F-R$) is well predicted by the onset of second mode of buckling $\bar{w}_2$. This is probably because motors restrict the first mode of buckling : in addition to walking on the filaments, motors are also elastic springs. If they are stiff, motors could act as pins (e.g. enforcing $y=0$ locally). If they are soft, they could act as a confinement along $y$ ; lateral confinement indeed tends to favor higher modes of buckling \cite{dmitrieff2017balance}.

While additional pinning can be seen as directly selecting higher modes of deformation, lateral confinement would subtly change the mode shape by changing the action $\mathcal{L}$ as : 
\begin{eqnarray}
\mathcal{L} = \frac{1}{2} \left(  \kappa \left( y'' \right)^2   + \sigma \left( y' \right)^2  + k y^2 \right) 
\end{eqnarray}
Unfortunately, this yields a full fourth order differential equation on $y(x)$ that we cannot solve analytically. 

In any case, the elasticity of motors will increase the buckling force at the pinned end ; we found that this buckling force is at least 2.6 times higher than the Euler Buckling force. Phenomenologically, it seems that the second mode of deformation, predicting a pinned-end buckling force $f_2^* \approx 9.7 \times f_B$, is a good candidate.



\end{document}